\documentclass[]{raa}            
\usepackage{graphicx,times}
\usepackage{amssymb}

\providecommand{\bm}[1]{\mbox{\boldmath$#1$\unboldmath}}

\begin{document}

   \title{Numerical study of self-gravitating protoplanetary discs}

 \volnopage{ {\bf 2011} Vol.\ {\bf 0} No. {\bf XX}, 000--000}
   \setcounter{page}{1}

   \author{Kazem Faghei
      \inst{}
    }

   \institute{School of Physics, Damghan University, Damghan, Iran; {\it kfaghei@du.ac.ir}\\
\vs \no
   {\small Received [year] [month] [day]; accepted [year] [month] [day] }
}

\abstract{ In this paper, the effect of self-gravity on the protoplanetary discs is investigated. The mechanisms of angular momentum transport and energy dissipation are assumed to be the viscosity due to turbulence in the accretion disc. The energy equation is considered in situation that the released energy by viscosity dissipation is balanced with cooling processes. The viscosity is obtained by equality of dissipation and cooling functions, and is used for angular momentum equation. The cooling rate of the flow is calculated by a prescription, $d u/d t=-u/\tau_{cool}$, that $u$ and $\tau_{cool}$ are internal energy and cooling timescale, respectively. The ratio of local cooling to dynamical timescales $\Omega \tau_{cool}$ is assumed as a constant and also as a function of local temperature. The solutions for protoplanetary discs show that in situation of $\Omega \tau_{cool} = constant$, the disc does not show any gravitational instability in small radii for a typically mass accretion rate, $\dot{M} = 10^{-6} M_{\odot} yr^{-1}$, while by choosing $\Omega \tau_{cool}$ as a function of temperature, the gravitational instability  for this amount of mass accretion rate or even less can occur in small radii. Also, by study of the viscous parameter $\alpha$, we find that the strength of turbulence in the inner part of self-gravitating protoplanetary discs is very low. These results are qualitatively consistent with direct numerical simulations of protoplanetary discs. Also, in the case of cooling with temperature dependence, the effect of physical parameters on the structure of the disc  is investigated. The solutions represent that disc thickness and Toomre parameter decrease by adding the ratio of disc mass to central object mass. While, the disc thickness and Toomre parameter increase by adding mass accretion rate. Furthermore, for typically input parameters such as mass accretion rate $10^{-6} M_{\odot} yr^{-1}$, the ratio of the specific heats  $\gamma=5/3$, and the ratio of disc mass to central object mass $q=0.1$, the gravitational instability can occur in whole radii of the discs  excluding very near to the central object. 
\keywords{accretion, accretion discs --- planetary systems: protoplanetary discs --- planetary systems: formation
}
}

   \authorrunning{K. Faghei}            
   \titlerunning{Self-gravitating protoplanetary discs}  
   \maketitle


%
%
\section{Introduction}
Accretion discs are important for many astrophysical phenomena, including protoplanetary systems, different types of binary stars, binary X-ray sources, quasars, and Active Galactic Nuclei (AGN). Historically, theory of accretion discs had concentrated in the case of non self-gravitating and occasionally the effect of self-gravity had studied (Paczy\'{n}ski 1978; Kolykhalov \& Sunyaev 1979; Lin \& Pringle 1987, 1990). On the other hand, in recent years, the importance of study of disc self-gravity has increased, especially in the protostellar discs and
Active Galactic Nuclei (AGN) discs. It can be due to increase of computational resources in simulation of self-gravitating accretion discs and 
the observational evidences that have confirmed the existence of self-gravity on all scale discs, from AGN to protostars (Lodato 2008 and references therein). Also, it appears the development of gravitational instability is important for cool regions of accreting gas that angular momentum transport by magneto-rotational instability (MRI) becomes weak (Fleming et al. 2000; Masada \& Sano 2008; Faghei 2011) and angular momentum can transport by gravitational instability.

The structure of self-gravitating discs has been studied both through self-similar solutions assuming steady and unsteady state (Mineshige \& Umemura 1996, 1997; Tsuribe 1999; Bertin \& Lodato 1999, 2001; Shadmehri \& Khajenabi 2006; Abbassi et. al. 2006; Shadmehri 2009) and through direct numerical simulations (Gammie 2001; Rice et al. 2003, 2005, 2010; Rice \& Armitage 2009; Cossins et al. 2010; Meru \& Bate 2011a).

Mineshige \& Umemura (1996) investigated the role of self-gravity on the classical self-similar solution of advection dominated accretion flows (ADAF, Narayan \& Yi 1994) and found global one-dimensional solutions influenced by self-gravity in both the radial and the perpendicular directions of the disc. They extended the previous steady state solutions to the time-dependent case while the effect of the self-gravity of the disc was taken into account. They used an isothermal equation, and so their solutions describe a viscous accretion discs in the slow accretion limit. Tsuribe (1999) studied unsteady viscous accretion in self-gravitating discs. Taking into account the growth of the central point mass, Tsuribe (1999) derived a series of self-similar solutions for rotating isothermal discs. The solutions showed, as a core mass increases, the rotation law changes from flat rotation to Keplerian rotation in the inner disc and in addition to the central point mass, the inner disc grows by mass accumulation due to the differing mass accretion rates in the inner and outer radii.  Bertin \& Lodato (1999) considered a class of steady-state self-gravitating accretion discs for which efficient cooling mechanisms are assumed to operate so that the disc is self-regulated at a condition of approximate marginal Jeans stability. They investigated the entire parameter space available for such self-regulated accretion discs. In another study, Bertin \& Lodato (2001) followed the model that, when the disc is sufficiently cold, the stirring due to Jeans-related instabilities acts as a source of effective heating. The corresponding reformulation of the energy equations, they demonstrated how self-regulation can be established, so that the stability parameter $Q$ is maintained close to a threshold value, with weak dependence on radius. Abbassi et al. (2006) studied the effect of viscosity on the time evolution of axisymmetric, polytropic self-gravitating discs around a new born central object. Thus, they ignored from the gravitational effect of central object and only self-gravity of the disc played an important role. They compared effects of $\alpha$-viscosity prescription (Shakura \& Sunyaev) and $\beta$-viscosity prescription (Duschel et al. 2000) on disc structure. They found that accretion rate onto the central object for $\beta$-discs more than $\alpha$-discs at least in the outer regions where $\beta$-discs are more efficient. Also, their results showed gravitational instability can occur everywhere on the $\beta$-discs and thus they suggested that $\beta$-discs can be a good candidate for the origin of planetary systems. Shadmehri \& Khejenabi (2006) examined steady self-similar solutions of isothermal self-gravitating discs in the presence of a global magnetic field. Similar to Abbassi et. al. (2006) they neglected from the mass of the central object to the disc mass. By study of Toomre parameter they showed that magnetic field can be important in gravitational stability of the disc.

An accretion discs can become gravitationally unstable if Toomre parameter becomes smaller than its critical value, $Q < Q_{crit}$ (Toomre 1964). For axisymmetric instabilities $Q_{crit} \sim 1$, while for non-axisymmetric instabilities $Q_{crit}$ values as high as $1.5\, -\, 1.7$ (Durisen et al. 2007). One possible outcome is that unstable discs fragment to produce bound objects and has been suggested as a possible mechanism for forming giant planets (Boss 1998, 2002). However, recently it has been realized that above condition is not sufficient to guarantee fragmentation. Gammie (2001) showed that in addition to the above instability criterion, the disc must cool at a fast enough rate. Let the cooling timescale $\tau_{cool}$ be defined as the gas internal energy divided by the volumetric cooling rate. For power-law equation of state and with $\tau_{cool}$ prescribed to be some value over a annulus of the disc, the thin shearing box simulations of Gammie (2001) show that fragmentation occurs if and only if $\Omega\tau_{cool} \lesssim \beta_{crit}$, where $\beta_{crit} \sim 3$ and $\Omega$ is angular velocity of the disc or inverse of dynamical timescale $\tau_{dyn}=\Omega^{-1}$. The critical value of $\Omega\tau_{cool}$ can be somewhat larger than three for more massive and physically thicker discs (Rice et al. 2003), larger adiabatic index (Rice et al. 2005), and more resolution of simulations (Meru \& Bate 2011b). Cossins et al. (2010) by SPH simulation studied the effects of opacity regimes on the stability of self-gravitating protoplanetary discs to fragmentation into bound objects. They showed that $\Omega\tau_{cool}$ has a strong dependence on the local temperature. As, they found that without temperature dependence, for radii $\lesssim 10 AU$ a very large accretion rate $~ 10^{-3} M_{\odot} yr^{-1}$ is required for fragmentation, but that this is reduced to $10^{-4}$ with cooling of dependent on temperature.

As mentioned, typically semi-analytical studies of self-gravitating discs are regarding polytropic discs (Abbassi et al. 2006), isothermal discs (Mineshige \& Umemura 1996, 1997; Tsuribe 1999; Shadmehri \& Khajenabi 2006), ADAFs in the extreme limit of no radiative cooling (Shadmehri 2004), and discs without central object (Mineshige \& Umemura 1996, 1997; Tsuribe 1999; Shadmehri \& Khajenabi 2006; Abbassi et al. 2006). In this paper, it will be interesting to understand under which conditions gravitational instability can occur in accretion discs by a suitable energy equation and assuming a Newtonian potential of a mass point that stands in the disc centre. Thus, to obtain these conditions,  we will use a prescription for cooling rate that is introduced by Gammie (2001),  $d u/d t=-u/\tau_{cool}$, that $u$ and $\tau_{cool}$ are internal energy and cooling timescale, respectively. The ratio of local cooling to dynamical timescales $\Omega \tau_{cool}$ is assumed a power-law function of temperature in adapting Cossins et al. (2010), $\Omega \tau_{cool}=\beta_0 (T/T_0)^\delta$, where $T_0$ and $\delta$ are free parameters, and $\beta_0$ is free parameter in Gammie (2001). In $\delta=0$, $\Omega \tau_{cool}$ reduces to Gammie (2001) model that $\Omega \tau_{cool}$ is a constant, while non-zero $\delta$ is qualitatively consistent with results of Cossins et al. (2010). We will examine the effects of $\delta$ parameter on gravitational stability of disc. We will show that the present model is qualitatively consistent with direct numerical simulations (Rice \& Armitage 2009; Cossin et al. 2010; Rice et al. 2010) and can provide conditions that gravitational instability occur in whole radii excluding very near to the central object.

In section 2, the basic equations of constructing a model for steady self-gravitating disc will be defined. In section 3, we will find asymptotic solutions for outer edge of the disc. In section 4, by exploit of asymptotic solutions as boundary conditions for system equations, we will investigate numerically the effects of physical parameters on structure and stability of the disc. The summary and discussion of the model will appear in section 5.

\section{Basic Equations}

We use cylindrical coordinate $(r,\varphi,z)$ centered on the accreting object and make the following standard assumptions:
\begin{itemize}
  \item [(i)] The flow is assumed to be steady and axisymmetric $ \partial_{t}=\partial_{\varphi}=0$, so all flow variables are a function of $r$ and $z$ ;
  \item [(ii)] The gravitational force of central object on a fluid element is characterized by the Newtonian potential of a point mass, $\Psi=-{GM_{*}}/{r}$,
  with $G$ representing the gravitational constant and $M_{*}$ standing for the mass of the central star; 
  \item [(iii)] The equations written in cylindrical coordinates are integrated in the vertical   direction, hence all quantities of the flow variables will be expressed in terms of cylindrical radius $r$;
  
\end{itemize}
The governing equations on the self-gravitating accretion disc for such assumptions is as follows. The continuity equation is 
\begin{equation}
\frac{1}{r}\frac{d}{d r}(r \Sigma v_{r})=0,
\end{equation}
where $v_r$ is the radial infall velocity and $\Sigma$ is the surface density, which is defined as $\Sigma=2 \rho h$, and $\rho$ and $h$ are density and the disk half-thickness, respectively. The half-thickness of the disc with assume of hydrostatic equilibrium in vertical direction is $ h=c_s/\Omega$, where $c_s$ is sound speed, which is defined as $c_s^{2}=p/\rho$, $p$ being the gas pressure and $\Omega$ represents angular velocity of the flow. The equation (1) implies that 
\begin{eqnarray}
\nonumber\dot {M}=-2 \pi r \Sigma v_{r}=constant
\end{eqnarray}
where $\dot{M}$ is the mass accretion rate and is a constant in the present model. The simulation results of protoplanetary discs show that the disc reaches a quasi-steady state in 20000 years or less and might imply that these systems are rarely out of equilibrium. Also, the simulations show that the mass of the disc redistribute itself to produce a state in which the accretion rate, $\dot{M}$, is largely independent of $r$ (Rice \& Armitage 2009, Rice et al. 2010). Thus, we can use the mass accretion as a constant and it can not be a limitation for the present model. The momentum equations are
\begin{equation}
v_{r}\frac{d v_{r}}{d r}=
  -\frac{1}{\Sigma}\frac{d }{d r}(\Sigma c_{s}^{2})
  -G\left[\frac{M_{*}+M(r)}{r^{2}}\right]
   +r\Omega^{2},
\end{equation}

\begin{equation}
\Sigma v_{r}\frac{d}{d r}(r^{2}\Omega)=
  \frac{1}{r}\frac{d}{d r}\left[\nu \Sigma r^{3}\frac{d \Omega}{\partial r}\right],
\end{equation}
where $\nu$ is the kinematic viscosity coefficient, and $\gamma$ is the adiabatic index, and $M(r)$ is the mass of a disc within a radius $r$. 
As in Mineshige \& Umemura (1997), we adopt the monopole approximation for the radial gravitational force due to the self-gravity of the disc, which considerably simplifies the calculations and is not expected to introduce any significant error as long as the surface density profile is steeper than $1/r$ (e.g. Li \& Shu 1997; Saigo \& Hanawa 1998; Tsuribe 1999; Krasnopolsky \& Konigl 2002; Shadmehri 2009). Now, we can write
\begin{equation}
\frac{d M(r)}{d r}=2 \pi r \Sigma.
\end{equation}
The energy equation is
\begin{equation}
 \frac{\Sigma v_r}{\gamma-1}\frac{d c_s^{2}}{d r}+
\frac{\Sigma c_s^{2}}{r}\frac{d}{d
r}\left( r v_r\right)= \Gamma-\Lambda,
\end{equation}
where $\Gamma$ is the heating rate of the gas by dissipation processes such as turbulent viscosity and $\Lambda$ represents the energy loss through radiative cooling processes. The forms of the dissipation and cooling functions can be written as
\begin{equation}
    \Gamma=r^2 \Sigma \nu |\frac{d \Omega}{d r}|^2
\end{equation}

\begin{equation}
    \Lambda=\frac{1}{\gamma (\gamma-1)}\frac{\Sigma c_s^2}{\tau_{cool}}
\end{equation}
where $\tau_{cool}$ is cooling timescale. As noted in the introduction, we are interest to consider the effect of cooling function on the structure of self-gravitating discs. Thus, similar to Rice \& Armitage (2009) we will study the effects of it in the case of the heating rate in disc is equal to cooling rate, $\Gamma=\Lambda$. 

Since fragmentation requires fast cooling, Gammie (2001) suggested the cooling timescale can be parameterized as $\beta=\Omega \tau_{cool}$ , where $\beta$ is a free parameter. Gammie (2001) showed fragmentation requires $\beta~\lesssim~\beta_{crit}$, where $\beta_{crit} \approx 3$ 
for the adiabatic index of $\gamma=2$. Rice et al. (2005) performed  3D simulations to show the dependence of $\beta_{crit}$ on $\gamma$: for discs with $\gamma=5/3$ and $7/5$, $\beta_{crit} \approx 6 - 7$ and $\approx 12 - 13$, respectively. Recently,  Cossins et al. (2010) studied $\beta$ as a function of temperature. They showed that $\beta$ has a strong dependence on the local temperature. They found that without temperature dependence, for radii $ \lesssim~10 au$ a very large accretion rate $~ 10^{-3}~M_\odot~yr^{-1}$ is required for fragmentation,
but that this is reduced to $10^{-4}~M_\odot~yr^{-1}$ with cooling of dependent on temperature. So, for simplicity in this paper  we will use a cooling timescale with a power-law dependence on temperature  for study of the equations (1)-(5) 
\begin{eqnarray}
\nonumber \tau_{cool} = \frac{\beta_0}{\Omega } (\frac{T}{T_0})^\delta \\
 =\frac{\beta_0}{\Omega} (\frac{c_s}{c_{s_0}})^{2\delta}
\end{eqnarray}
that 
$\delta$ and $\beta_0$ are free parameters, and if we select $T_0$ as a temperature of the outer part of the disc, then $c_{s_0}$ will be sound speed in there. From equation (8) and $\delta = 0$, we expect that $\Omega \tau_{cool}$ becomes a constant that is same with Gammie (2001) model. While non-zero $\delta$ is qualitatively consistent with Cossins et al. (2010) model. It is important to stress that the above description for cooling rate is not meant to reproduce any specific cooling law, but is just a convenient way of exploring the role of the cooling timescale in the outcome of the gravitational instability.

Here, the kinematic coefficient of viscosity can be obtained by equating of the heating and cooling rates
\begin{equation}
    \nu=\frac{1}{\gamma (\gamma-1)} 
\frac{ \left|\frac{d \Omega}{d r}\right|^{-2}}{r^2}
\frac{c_s^2}{\tau_{cool}}.
\end{equation}
Thus, by exploit of equation (9) we do not need to use of viscosity descriptions, such as $\alpha$ and $\beta$ prescriptions that are introduced by Shakura \& Sunyaev (1973) and Duschel et al. (2000), respectively. Equation (9) implies that the kinematic coefficient of viscosity in the present model depends on physical quantities of the system, specially cooling timescale. The kinematic coefficient of viscosity in $\alpha$-prescription  is $\nu=\alpha c_s h$, where $\alpha$ is a free parameter and is less than unity (Shakura \& Sunyaev 1973).
By using equation (9) for $\alpha$ parameter we can write
\begin{eqnarray}
    \nonumber\alpha=\frac{\nu}{c_s h}~~~~~~~~~~~~~~~~~~~~~~~\\
 =\frac{1}{\gamma (\gamma-1)} 
\frac{ \left|\frac{d \Omega}{d r}\right|^{-2}}{r^2 h}
\frac{c_s}{\tau_{cool}}.
\end{eqnarray} 
The above equation implies that the $\alpha$ parameter is not a constant and varies by position and strongly depends on cooling timescale. We will study the $\alpha$ parameter in section 4 and will show that in the present model it increases by radii.  

As mentioned in the introduction, the gravitational stability of the disc can be investigated by Toomre parameter (Toomre 1964). The Toomre parameter for an epicyclic motion can be written as
  \begin{equation}
   Q=\frac{c_s k}{\pi G \Sigma}
  \end{equation}
where
 \begin{equation}
  k=\Omega\sqrt{ 4+2 \frac{d \log \Omega }{d \log r}}
 \end{equation}
is the epicyclic frequency which can be replaced by the angular frequency, $\Omega$. 

The equations of (1)-(5) and (9) provide a set of ordinary differential equations that describes physical properties of the self-gravitating disc. Since, these equations are nonlinear, we will need suitable boundary conditions to solve it numerically. Thus, in next section we will try to obtain asymptotic solution in outer edge of the disc and then by exploit of this asymptotic solution as a boundary condition, we will able to integrate system equations inward from a point very near to the outer edge of the disc.

Before next sections and the numerical study of the model, we shall express all quantities in units with values typical protostellar disc. We will choose astronomical unit ($au$) and the sun mass ($M_{\odot}$) as the units of length and mass, respectively. Thus, the time unit is given by $\sqrt{au^3/G M_{\odot}}$ that is equal to a year divided to $2 \pi$.

\section{Outer Limit}
Here, the asymptotic behavior of the system equations as $r \rightarrow R$ is investigated that $R$ is the outer radius of the disc. The asymptotic solutions are given by

\begin{equation}
   \Sigma(r) \sim \frac{\Sigma_0}{ R^{1/2}}~\left(1+a_1 \frac{s}{R} + \cdot\cdot\cdot\right)  
\end{equation}

\begin{equation}
   v_r(r) \sim -c_1 \sqrt{\frac{M_{*}+M_{disc}}{R}}~\left(1+a_2 \frac{s}{R} + \cdot\cdot\cdot\right)  
\end{equation}

\begin{equation}
   \Omega(r) \sim c_2 \sqrt{\frac{M_{*}+M_{disc}}{R^3}}~\left(1+a_3 \frac{s}{R} + \cdot\cdot\cdot\right)  
\end{equation}

\begin{equation}
   c_s^2(r) \sim c_3 \frac{ M_{*}+M_{disc}}{R}~\left(1+a_4 \frac{s}{R} + \cdot\cdot\cdot\right)  
\end{equation}

\begin{equation}
   M(r) \sim M_{disc} - \int^R_{r} 2 \pi r^\prime \Sigma(r^\prime) dr^\prime 
\end{equation}
where $s=R-r$, $M_{disc}$ is the disc mass, and the coefficients of $c_i$, $a_i$, and $\Sigma_0$ must be determined. Using these solutions, 
from the continuity, momentum, angular momentum, energy, and viscosity  equations [(1)-(5), and (9)], we can obtain the coefficients of $c_i$ that have the 
following forms:
\begin{equation}
  c_1= \frac{\dot{M}}{2\pi\Sigma_0\sqrt{M_*+M_{disc}}}
\end{equation}

\begin{eqnarray}
\nonumber c_2^2+\left[\frac{a_3 \gamma (\gamma-1) \beta_0 \dot{M} (a_3-2) (a_1+a_4)   }{2 \pi \Sigma_0 \sqrt{M_*+M_{disc}}(a_1+a_3+a_4-1)}\right]c_2 \\
+\left[ \frac{a_2 \dot{M}^2}{4 \pi^2 \Sigma_0^2 (M_*+M_{disc})} -1 \right]=0
\end{eqnarray}

 \begin{equation}
  c_3= \left(\frac{ a_3 \gamma \beta_0 (a_3-2) (\gamma-1) 
\dot{M} }{2\pi\Sigma_0 (a_1+a_3+a_4-1)\sqrt{M_*+M_{disc}}} \right) c_2 
\end{equation}
where 
\begin{equation}
   a_4=(1+a_2)(1-\gamma). 
\end{equation}
The amount of mass accretion rate can be determined by observational evidences of the protoplanetary discs. Also, $\Sigma_0$ approximately can be determined by disc mass, $M_{disc}  \sim \pi R^2 \Sigma $. Thus, knowing the amounts of $\Sigma_0$ and $\dot{M}$ from the observations, the value of $c_3$ coefficient is only depended on value of $c_2$. On the other hand, the value of $c_2$ can be obtained by equation (19). Since, we have only one equation for coefficients of $a_i$ (equation 21), we will select below values for them in duration of numerical integration of system equations to obtain physical results
\begin{equation}
a_1 < -2 +\frac{3}{2}\, \gamma, ~~ 3\, a_2 = a_3 = \frac{3}{2}, ~~ a_4 = (1+a_2)(1-\gamma).
\end{equation}

\input{epsf}
\begin{figure}
\begin{center}
\centerline
{ 
{\epsfxsize=5.0cm\epsffile{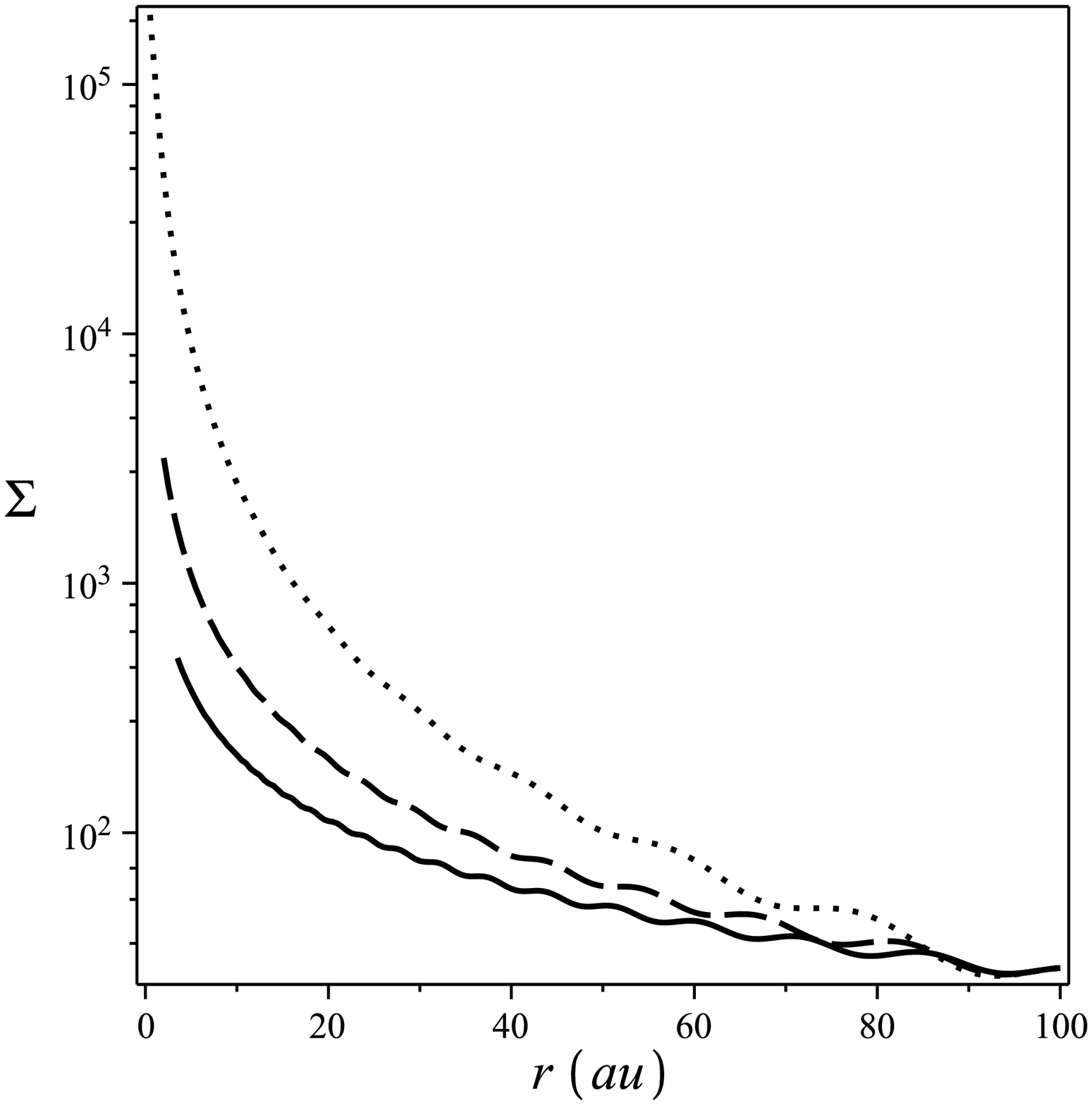}  }{\epsfxsize=5.0cm\epsffile{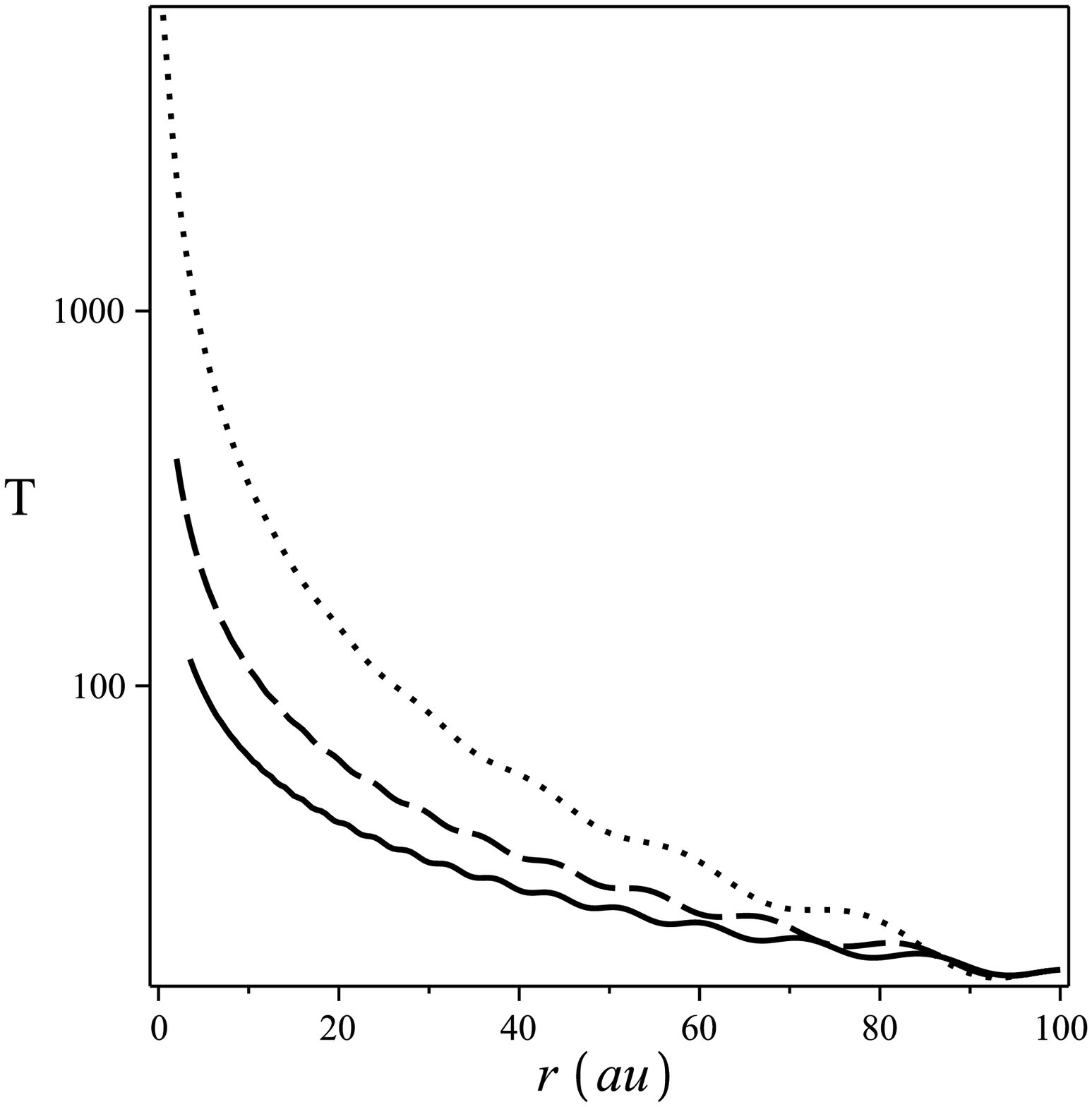}  }
} 
\centerline
{ 
{\epsfxsize=5.0cm\epsffile{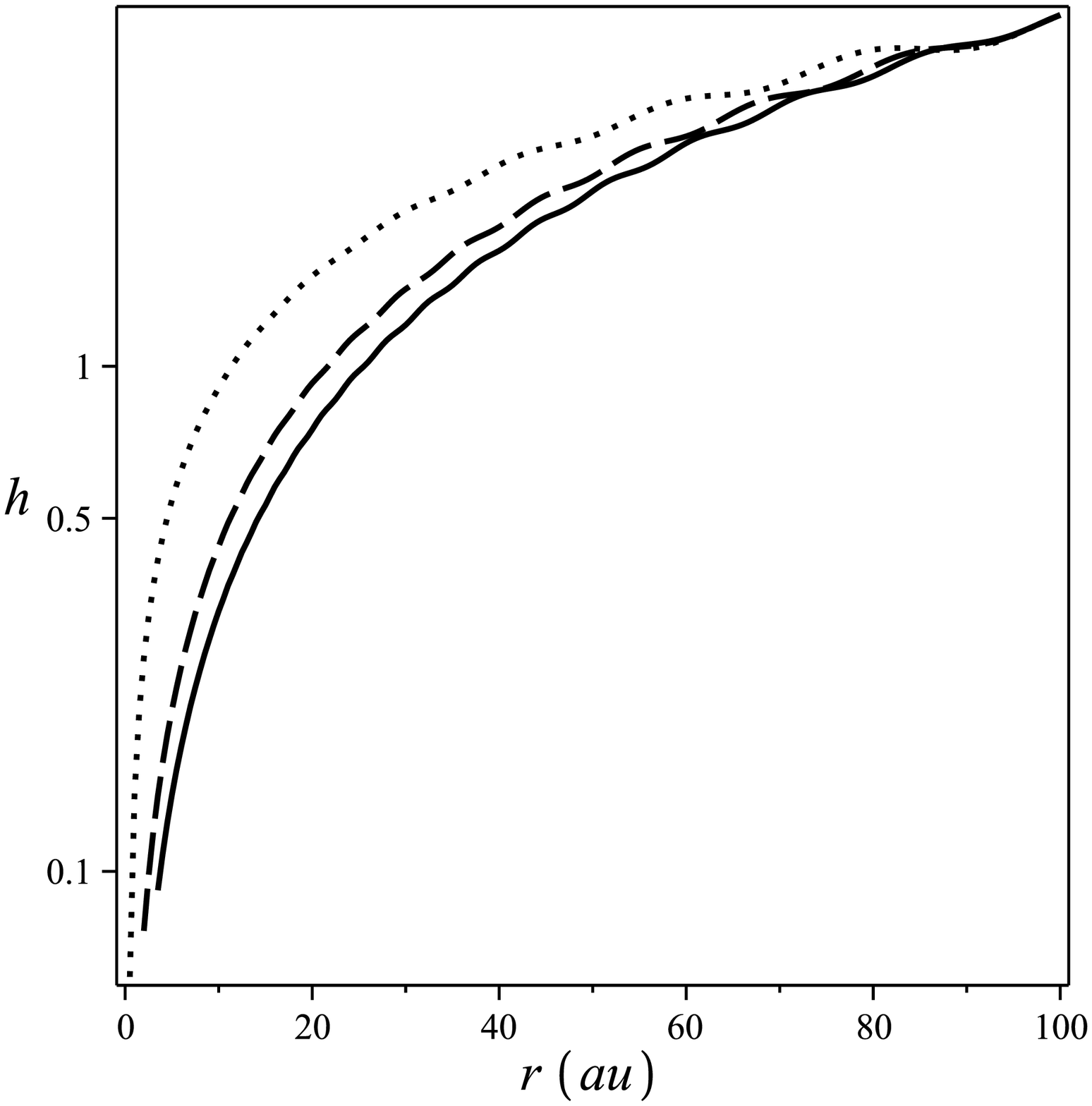}  }{\epsfxsize=5.0cm\epsffile{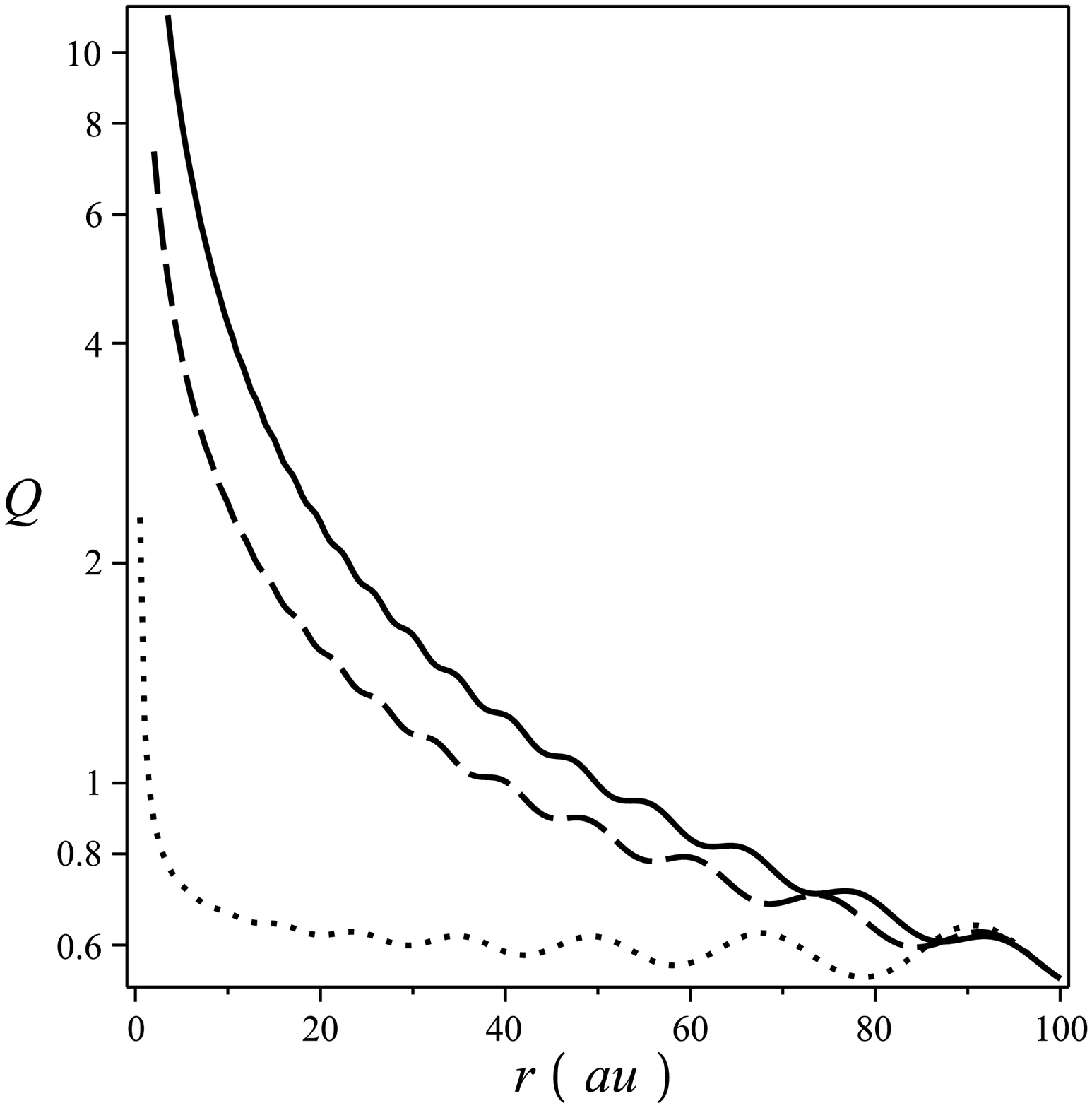}  }
}  
\end{center}
\vspace{-0.55cm}
\begin{center}
\caption{Surface density, thickness, temperature, and Toomre parameter of the disc as a function of radius, for several values of $\delta$. The surface density and the temperature are in $cgs$ system, and the thickness and the distance are in $au$ unit.  The solid lines represent $\delta=0$, the dashed lines represent $\delta=0.75$, and the dotted lines represent $\delta=1.5$. The input parameters are set to
the disc mass $M_{disc}=0.1 M_{\odot}$, the star mass $M_*=M_{\odot}$, the mass accretion rate  $\dot{M}=10^{-6} M_{\odot} yr^{-1}$, the ratio of the specific heats is set to be $\gamma=5/3$, and $\beta_0=2$. }
\end{center}
\end{figure}

\section{Numerical Results}
If the value of $R$ is guessed, the equations by Fehlberg-Runge-Kutta fourth-fifth order method can be integrated inwards from a point very near to the outer edge of the disc, using the above expansions. Examples of such solutions for surface density, half-thickness of the disc, temperature, Toomre parameter, and the viscous parameter of $\alpha$ as a function of radius are presented in Figs 1-5. The delineated quantities of $T$ in Figs 1-4 is the mid-plane temperature and can then be determined using
\begin{eqnarray}
\nonumber T=\left(\frac{\mu m_p}{k_B}\right)c_s^2
\end{eqnarray}
where $\mu=2$ is the mean molecular weight, $m_p$ is the proton mass, and $k_B$ is Boltzmann's constant.

\subsection{The influences of physical parameters on the results}
The free parameters in the present model are the importance degree of temperature in cooling timescale, $\delta$, the mass accretion rate, $\dot{M}$, the parameter of $\beta_0$, the ratio of disc mass to star mass, $q=M_{disc}/M_*$.

\subsubsection{$\delta$ parameter} 
The effects of $\delta$ parameter on the physical quantities are presented in Fig. 1. The profiles of surface density and temperature show that
they increase by adding $\delta$. But, the increase of surface density is more than temperature. Thus, the Toomre parameter ($Q \propto c_s/\Sigma \propto \sqrt{T}/\Sigma$) decreases by adding $\delta$ parameter. The profiles of Toomre parameter represent that for small $\delta$, only outer part of the disc gravitationally is unstable, and the gravitational instability can extend to inner radii by adding $\delta$ parameter. For $\delta_{crit} \sim 1.5$, the Toomre parameter in radii $\gtrsim 5\, au$ becomes smaller than critical Toomre parameter  ($Q_{cri}\sim 1$) and the disc becomes gravitationally unstable. In the other words, the profiles of Toomre parameter  represent the gravitational instability of the flow strongly depends on cooling timescale with temperature dependence. This result is qualitatively consistent with direct numerical simulations (e. g. Cossins et al. 2010). The disc thickness increases by adding $\delta$ parameter. It can be due to the increase of the temperature ($h \propto c_s \propto \sqrt{T}$).
 
Equations 8 and 9 imply that
\begin{equation}
  \frac{\nu_{(\delta\neq 0)}} {\nu_{(\delta=0)}}=  \left( \frac{c_s}{c_{s_0}} \right)^{-2 \delta}.
\end{equation}
Since $c_s \geq c_{s_0}$ the right-hand side of above equation is equal or less than one. On the other hand, non-zero $\delta$ constrains
lower viscosity for hotter regions of the disc. The study of gravitational instability shows that it enhances in lower viscosity (Abbassi et al. 2006; Shadmehri \& Khajenabi 2006; Khajenabi \& Shadmehri 2007). Thus, the gravitational instability can be enhanced by adding the $\delta$ parameter for hotter regions. But there is a limitation for the amount of $\delta$ parameter that we discuss it in next section. 

\input{epsf}
\begin{figure}
\begin{center}
{ 
{\epsfxsize=5.0cm\epsffile{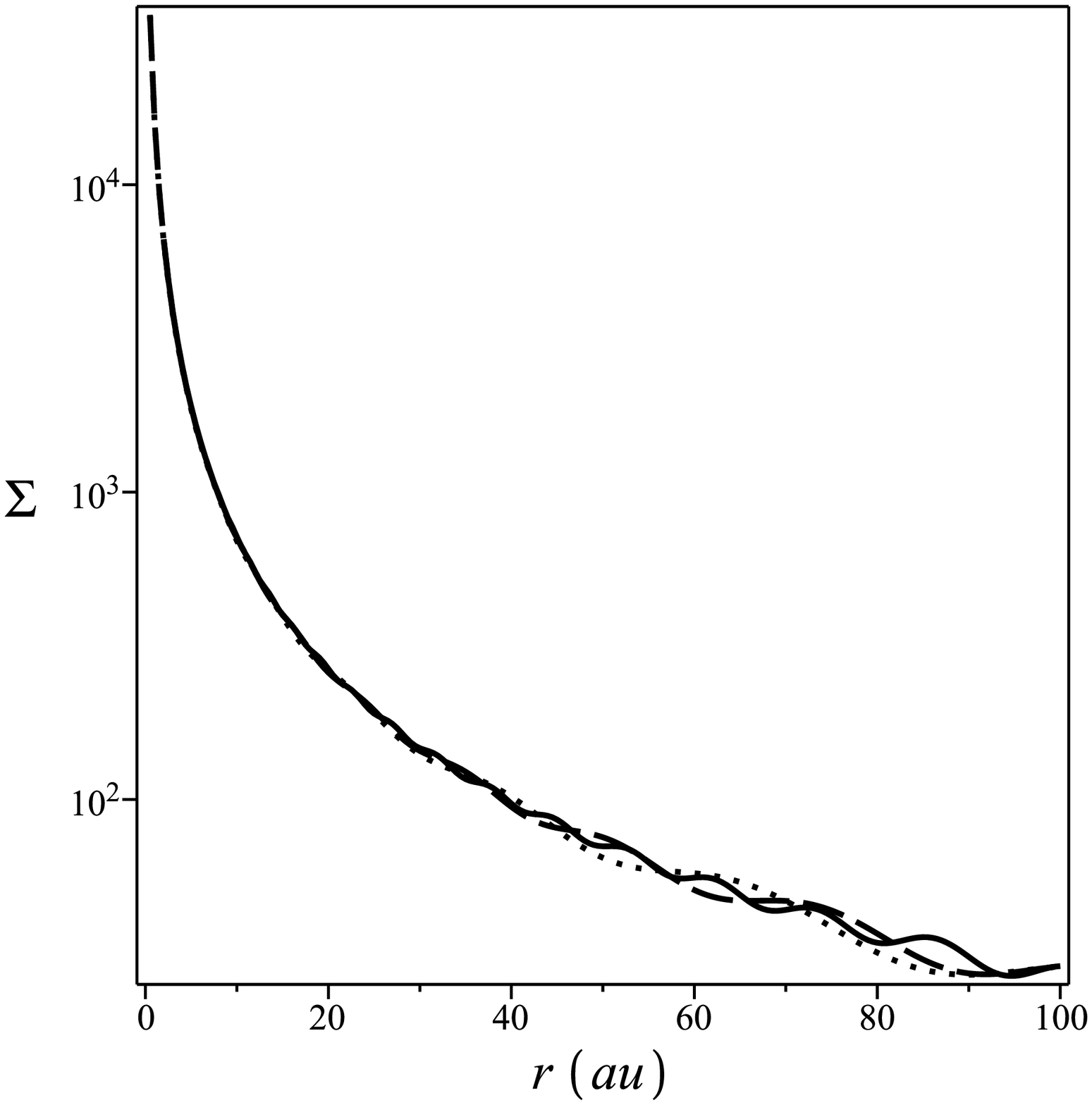}  }{\epsfxsize=5.0cm\epsffile{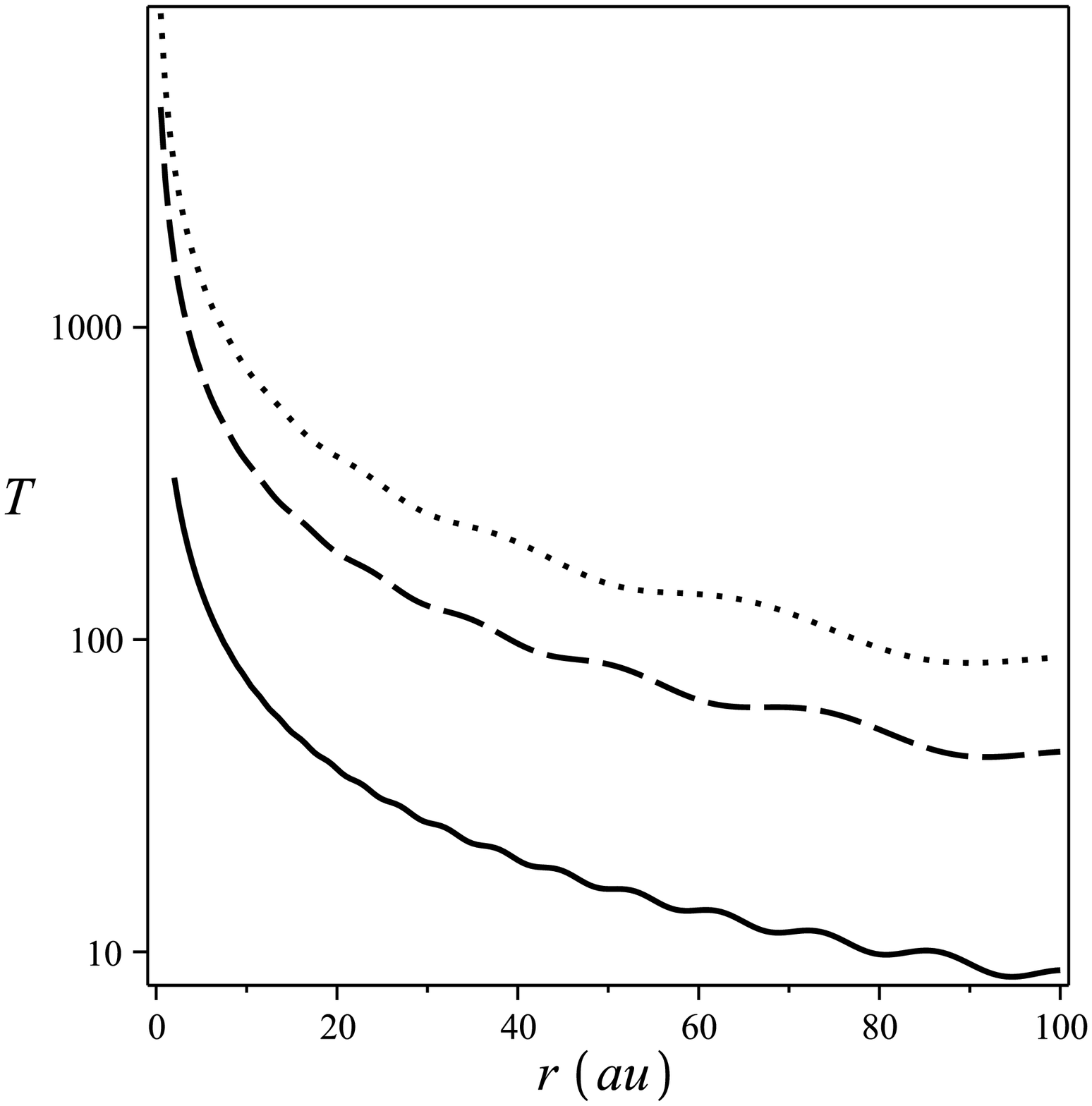}  }
} 
\centerline
{ 
{\epsfxsize=5.0cm\epsffile{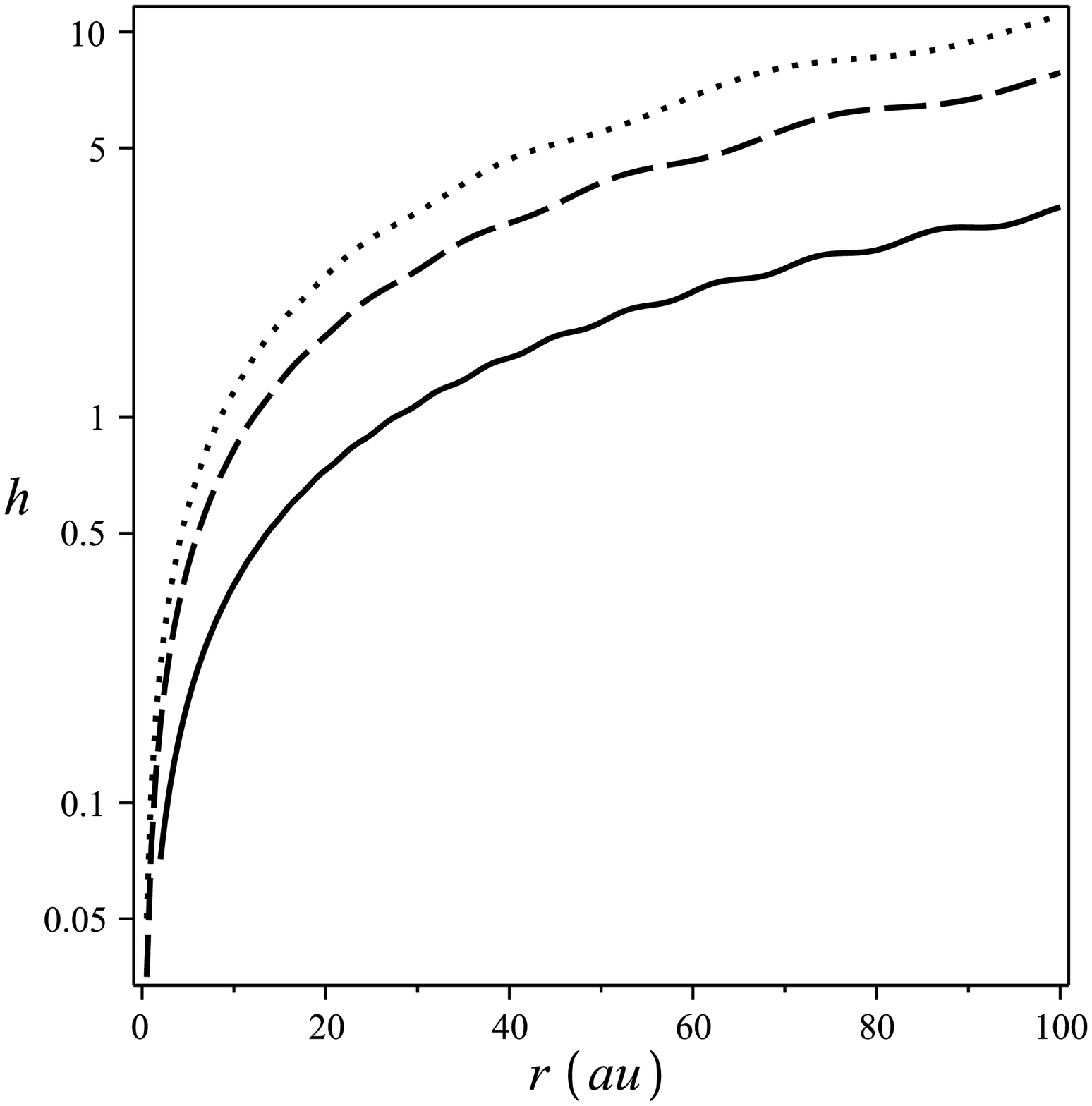}  }{\epsfxsize=5.0cm\epsffile{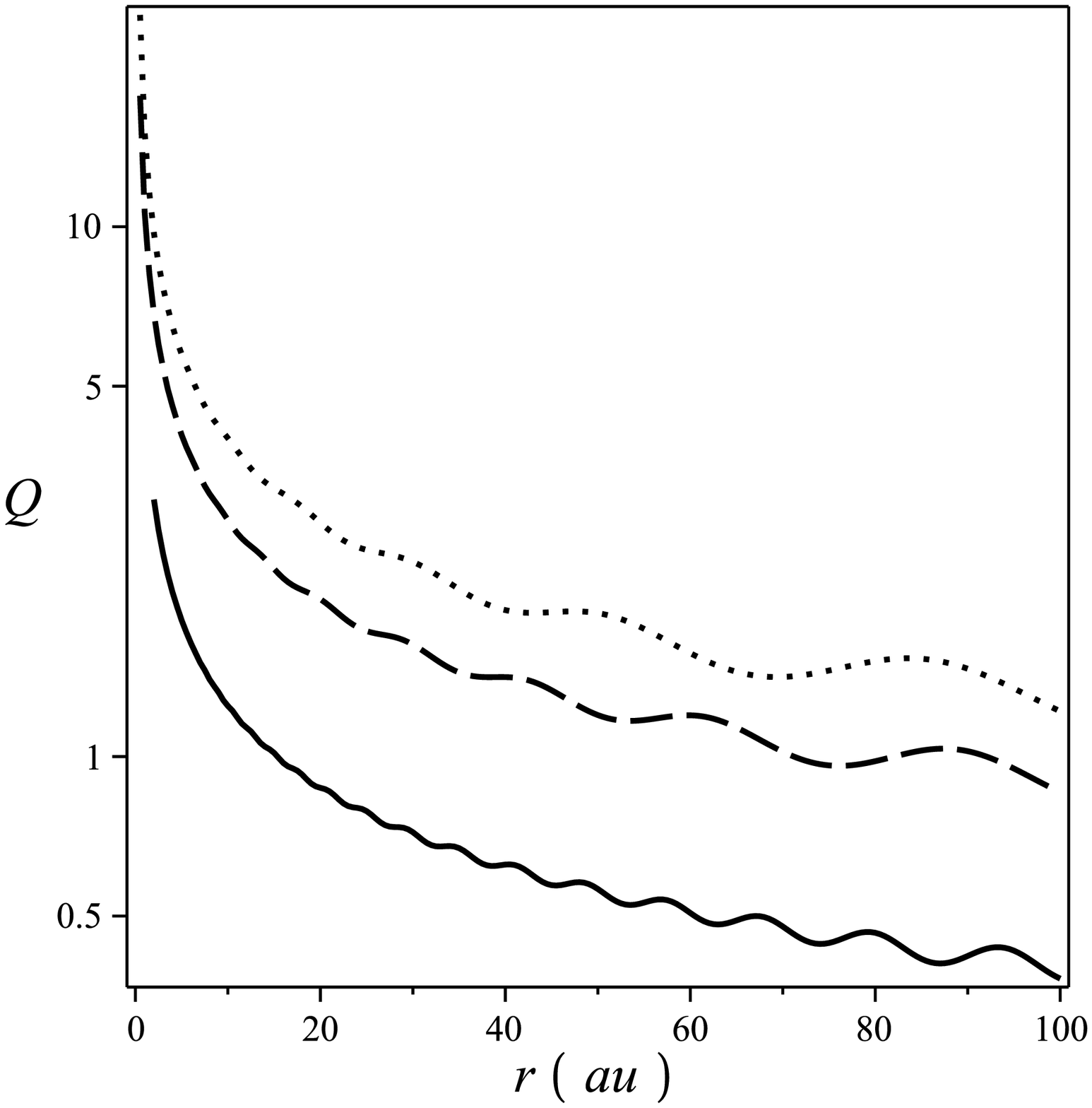}  }
}  
\end{center}

\begin{center}
\caption{Surface density, thickness, temperature, and Toomre parameter of the disc as a function of radius, for several values of $\beta_0$.  The surface density and the temperature are in $cgs$ system, and the thickness and the distance are in $au$ unit. The solid lines represent $\beta_0=1$, the dashed lines represent $\beta_0=5.0$, and the dotted lines represent $\beta_0=10$. The input parameters are set to the disc mass $M_{disc}=0.1 M_{\odot}$, the star mass $M_*=M_{\odot}$, the mass accretion rate $\dot{M}=10^{-6} M_{\odot} yr^{-1}$, the ratio of the specific heats is set to be $\gamma=5/3$, and $\delta=1.0$. }
\end{center}
\end{figure}

\subsubsection{$\beta_0$ parameter} 
The influences of  parameter of $\beta_0$ are shown in Fig. 2. As, we know from the simulations of self-gravitating disc (Gammie 2001; Rice et al. 2003), the reduce of this parameter provides conditions that the disc places on gravitational instability and consequently fragmentation. The profiles of surface density show that it does not change by adding the $\beta_0$ parameter and only it shows small deviations in large radii. The disc temperature increases by adding the $\beta_0$ parameter. Because, the increase of this parameter reduces the rate of cooling.
In large amount of $\beta_0$ ($\sim10$), the disc is gravitationally stable, while by reduce of its value to $5$, the gravitational instability can occur in large radii, and for the small value of it ($\beta_0\sim 1$), we can expect gravitational instability in whole of the disc excluding near to the star. These results are qualitatively consistent with direct numerical simulations of protoplanetary disc (Gammie 2001; Rice et al. 2003; Cossins et al. 2010). Also, the solutions show that the disc thickness increases by adding the $\beta_0$ parameter.

\input{epsf}
\begin{figure}
\begin{center}
{ 
{\epsfxsize=5.0cm\epsffile{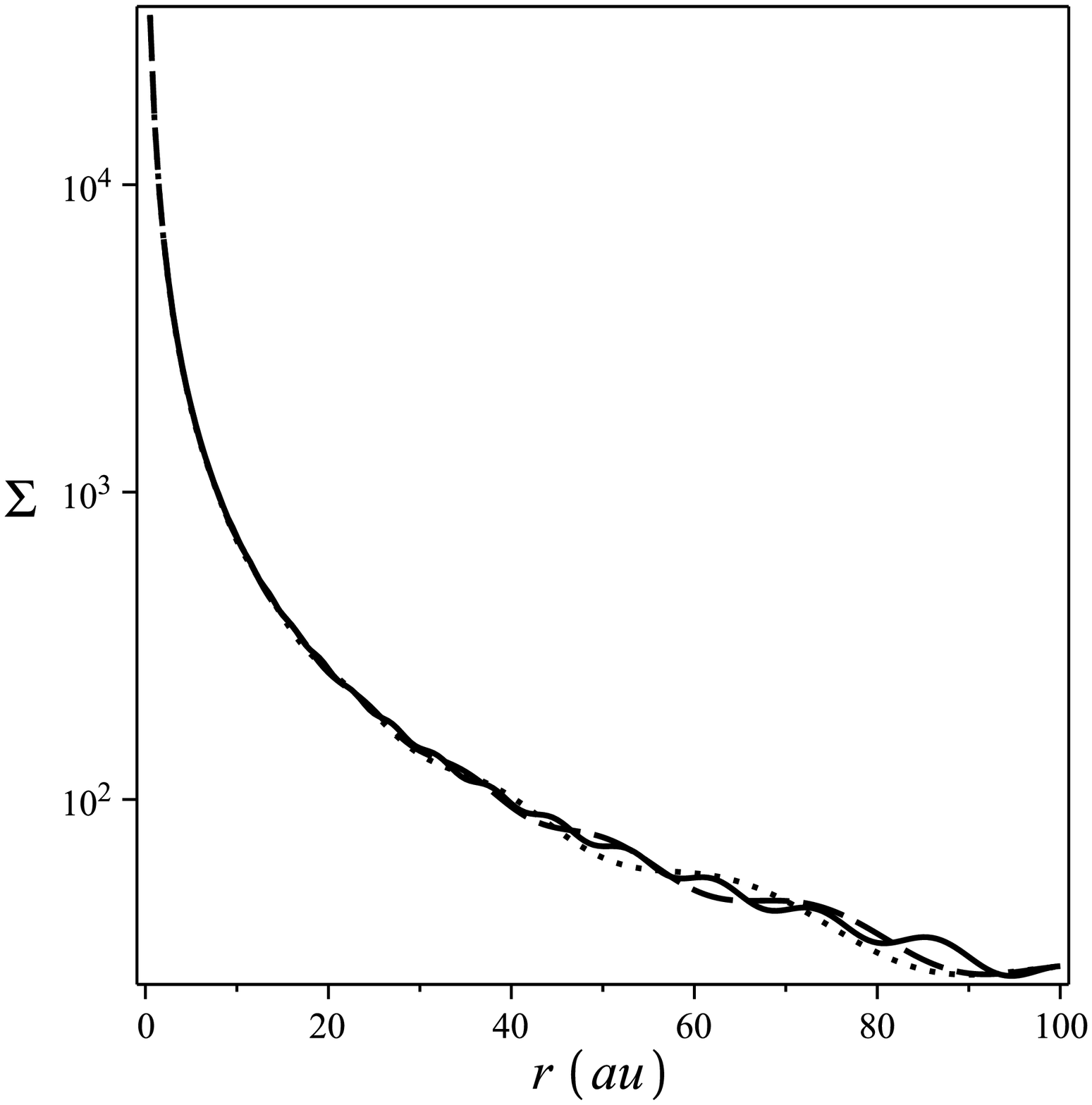}  }{\epsfxsize=5.0cm\epsffile{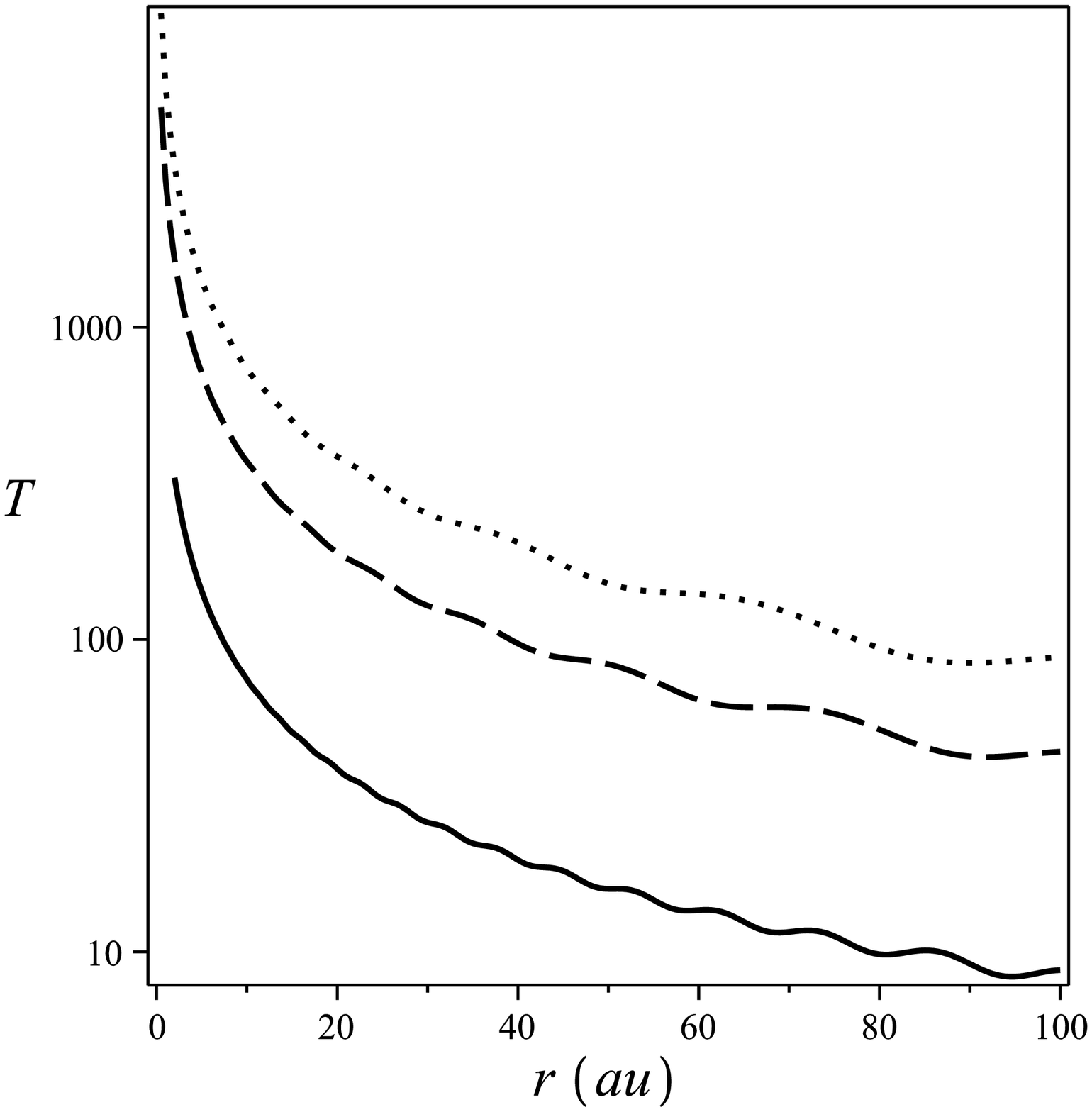}  }
} 
\centerline
{ 
{\epsfxsize=5.0cm\epsffile{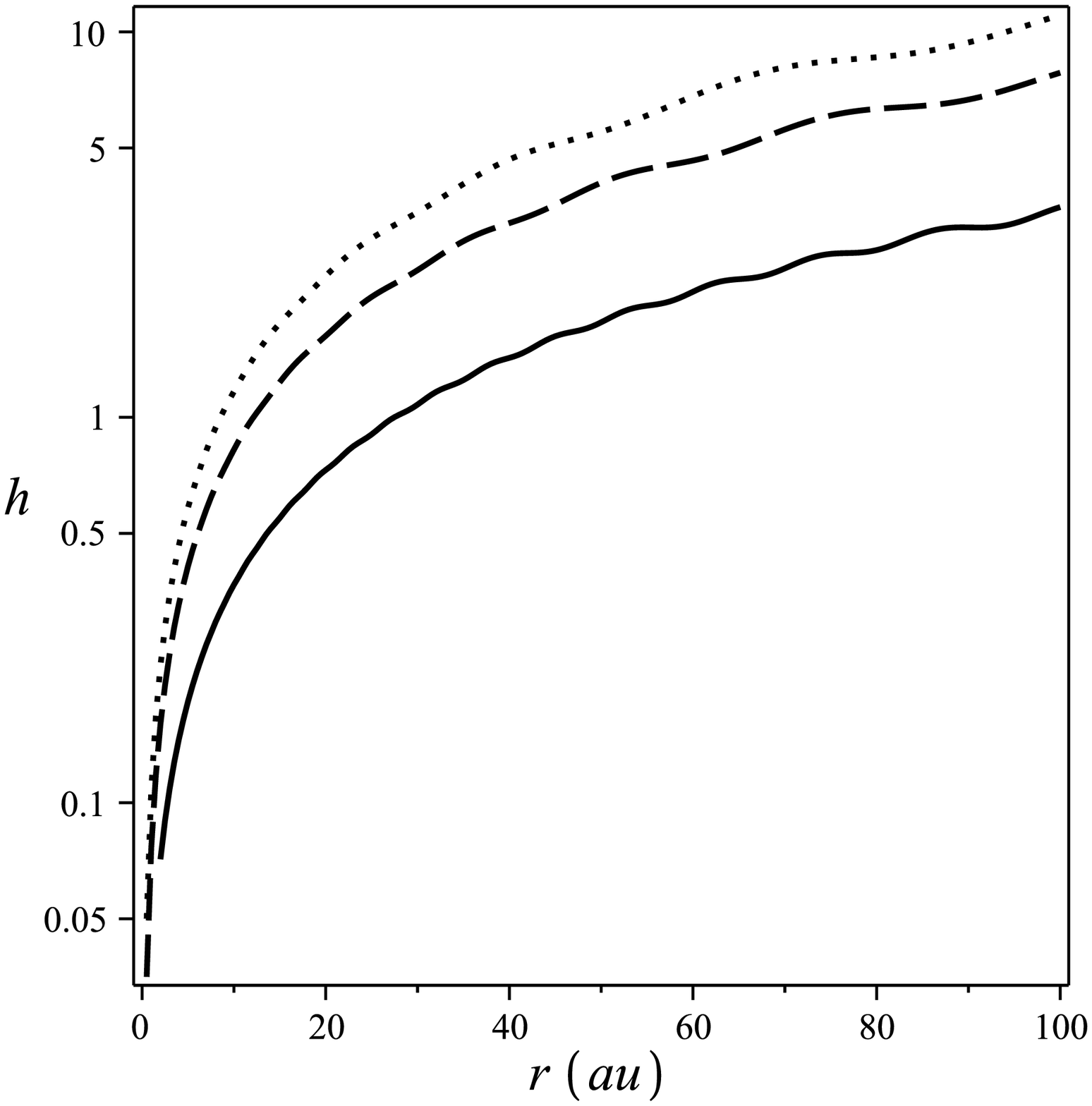}  }{\epsfxsize=5.0cm\epsffile{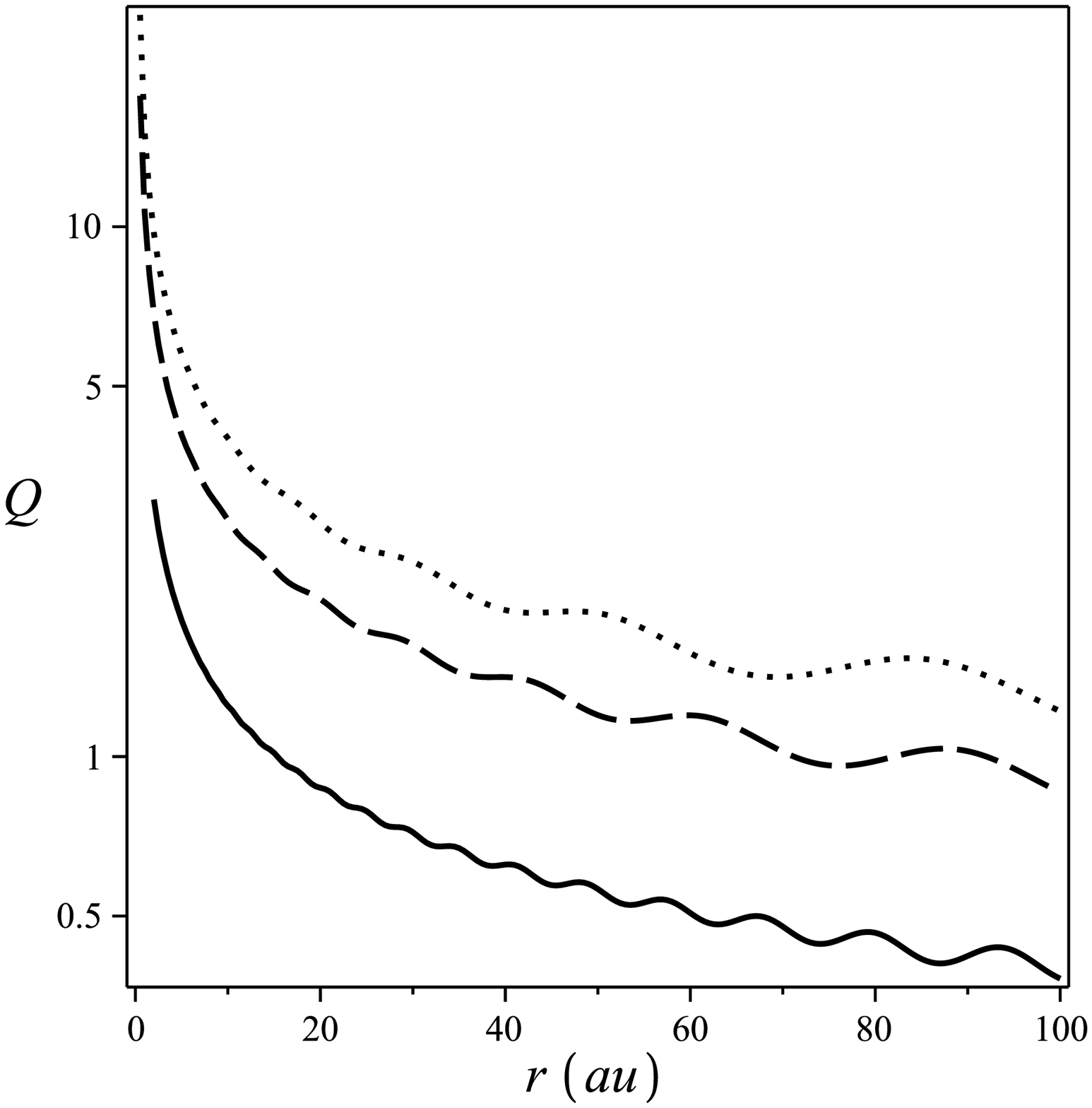}  }
}   
\end{center}

\begin{center}
\caption{Surface density, thickness, temperature, and Toomre parameter of the disc as a function of radius, for several values of $\dot{M}$. The surface density and the temperature are in $cgs$ system, and the thickness and the distance are in $au$ unit.  The solid lines represent $\dot{M}=10^{-7} M_{\odot} yr^{-1}$, the dashed lines represent $\dot{M}=5 \times 10^{-7} M_{\odot} yr^{-1}$, and the dotted lines represent $\dot{M}=10^{-6} M_{\odot} yr^{-1}$.  The input parameters are set to the disc mass $M_{disc}=0.1 M_{\odot}$, the star mass $M_*=M_{\odot}$, 
the ratio of the specific heats is set to be $\gamma=5/3$, $\beta_0=10$ and $\delta=1.0$. }
\end{center}
\end{figure}

\subsubsection{The mass accretion rate} 
Rice \& Armitage (2009) showed that beyond of $1\,au$ the disc reaches a quasi-steady state in $20000$ years and mass is redistributing itself to produce a state in which the accretion rate is largely independent of $r$. The mass accretion rate in their simulations finally reached to $10^{-6}-10^{-7} M_\odot/yr$ (see Fig 4 in their paper). We will study the behavior of the present model in Fig 3 for several values of the mass accretion rate ($10^{-7}$, $5 \times 10^{-7}$, and $10^{-6} M_\odot / yr$). The solutions imply that the disc temperature is sensitive to the amount of mass accretion rate and increases by adding the mass accretion rate. While, the surface density is not sensitive to the mass accretion rate and only shows small variations in large radii. Thus, the behavior of the temperature only specifies the behavior of the Toomre parameter ($Q \propto \sqrt{T}/\Sigma$). The profiles of Toomre parameter represent that it increases by adding the mass accretion rate. Also, the solutions show the disc thickness increases by adding mass accretion rate, that is due to increase of the disc temperature. The solutions show that for a low mass accretion rate ($\sim 10^{-7} M_\odot / yr$), but cooling timescale with temperature dependence ($\delta\sim 1$), the gravitational instability can occur for radii $\gtrsim 10\, au$.

\input{epsf}
\begin{figure}
\begin{center}
{ 
{\epsfxsize=5.0cm\epsffile{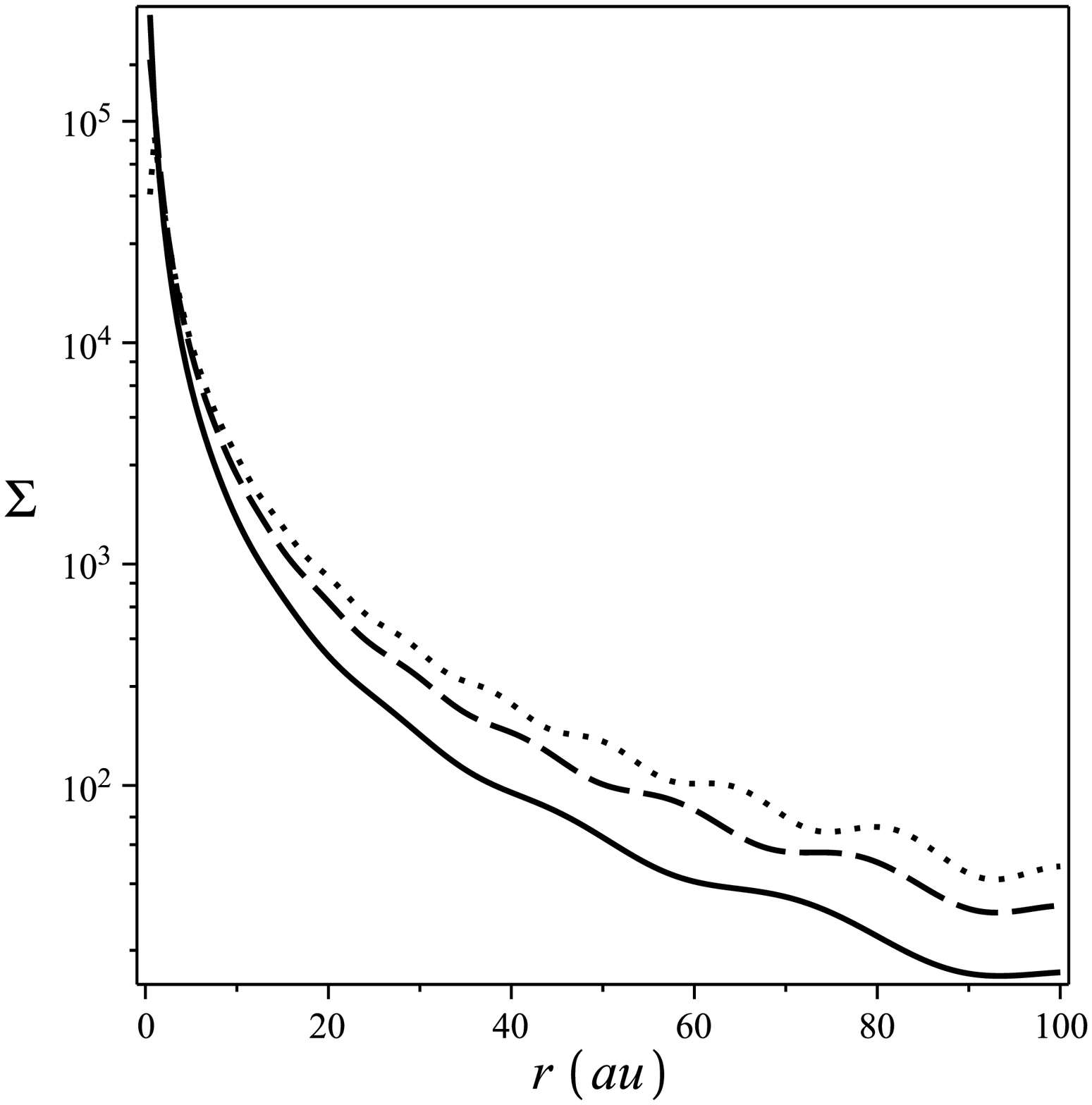}  }{\epsfxsize=5.0cm\epsffile{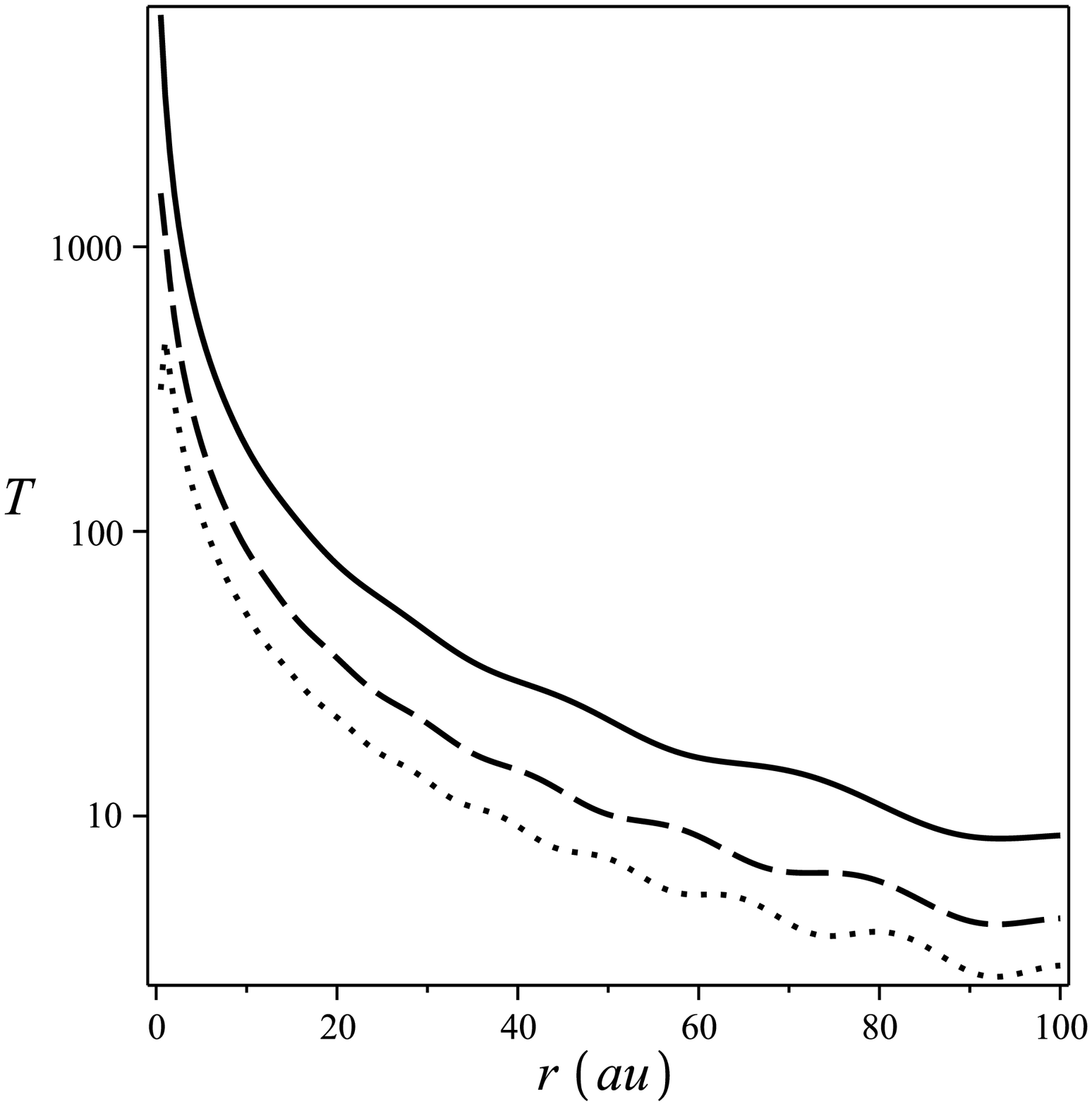}  }
} 
\centerline
{ 
{\epsfxsize=5.0cm\epsffile{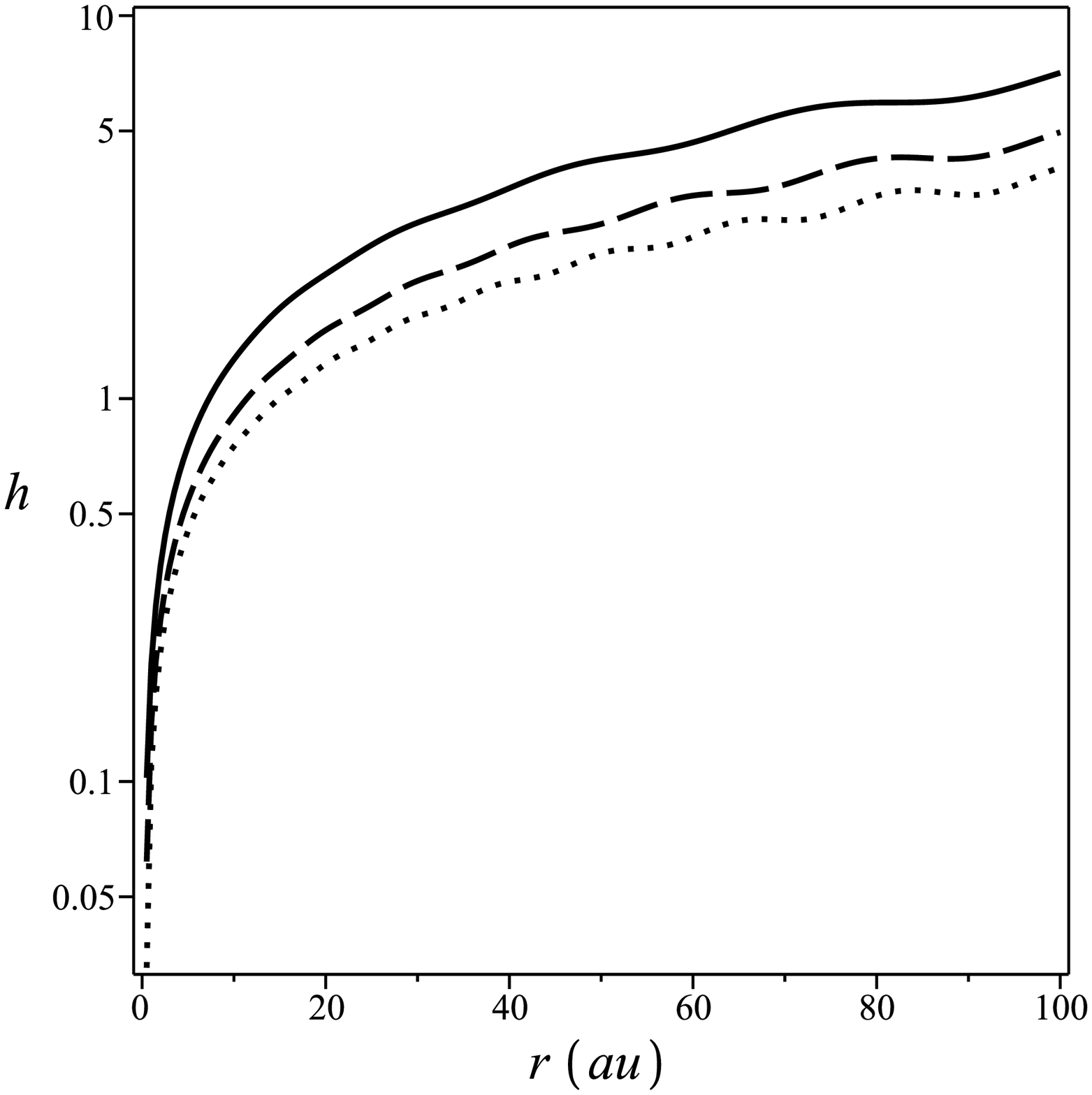}  }{\epsfxsize=5.0cm\epsffile{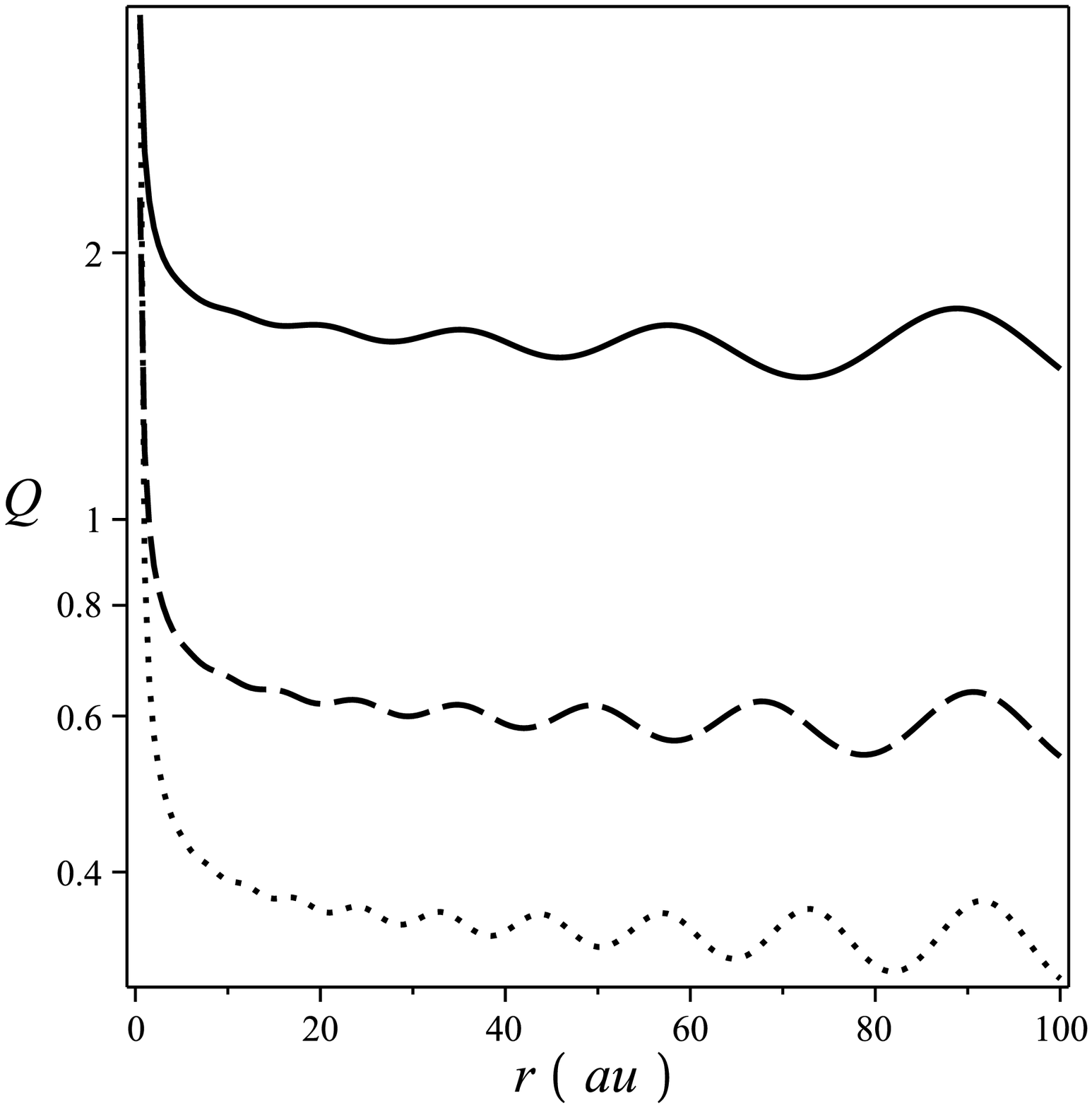}  }
}      
\end{center}

\begin{center}
\caption{Surface density, thickness, temperature, and Toomre parameter of the disc as a function of radius, for several values of $q=M_{disc}/M_{*}$. The surface density and the temperature are in $cgs$ system, and the thickness and the distance are in $au$ unit. The solid lines represent $q=0.05$, the dashed lines represent $q=0.1$, and the dotted lines represent $q=0.15$. The input parameters are set to the star mass $M_*=M_{\odot}$, the mass accretion rate  $\dot{M}=10^{-6} M_{\odot} yr^{-1}$, the ratio of the specific heats is set to be $\gamma=5/3$, $\beta_0=2$ and $\delta=1.5$.}
\end{center}
\end{figure}

\subsubsection{Mass ratio} 
As noted in the introduction, semi-analytical studies of self-gravitating discs are regarding discs without central object. This simplification is relevant to protostellar discs at the beginning of the accretion phase, during which the mass of the central object is small and only self-gravity of the disk plays an important role. Also, this simplification can correspond to discs at large radii because the effects of the central mass become unimportant in the outer regions of the disc. As, the central object attends in the present model and its effects are not ignored. Thus, the present model does not have limitations of previous studies of semi-analytical self-gravitating discs and can be applied for all region of the disc. Fig. 4 represents the effects of the ratio of the disc mass to the star mass $q=M_*/M_{disc}$ on the present model. 
The solutions show the surface density increases and the temperature decreases. Each of the surface density increasing and the temperature decreasing individually can reduce the Toomre parameter. Thus, we expect that Toomre parameter decreases by adding $q$ parameter that the profiles of Toommre parameter confirm it. The disc thickness profiles represent the disc thickness decreases by adding the disc mass. This property is qualitatively consistent with two-dimensional study of self-gravitating disc (e. g. Ghanbari \& Abbassi 2004). 

\input{epsf}
\begin{figure}
\centerline
{ 
{\epsfxsize=5.0cm\epsffile{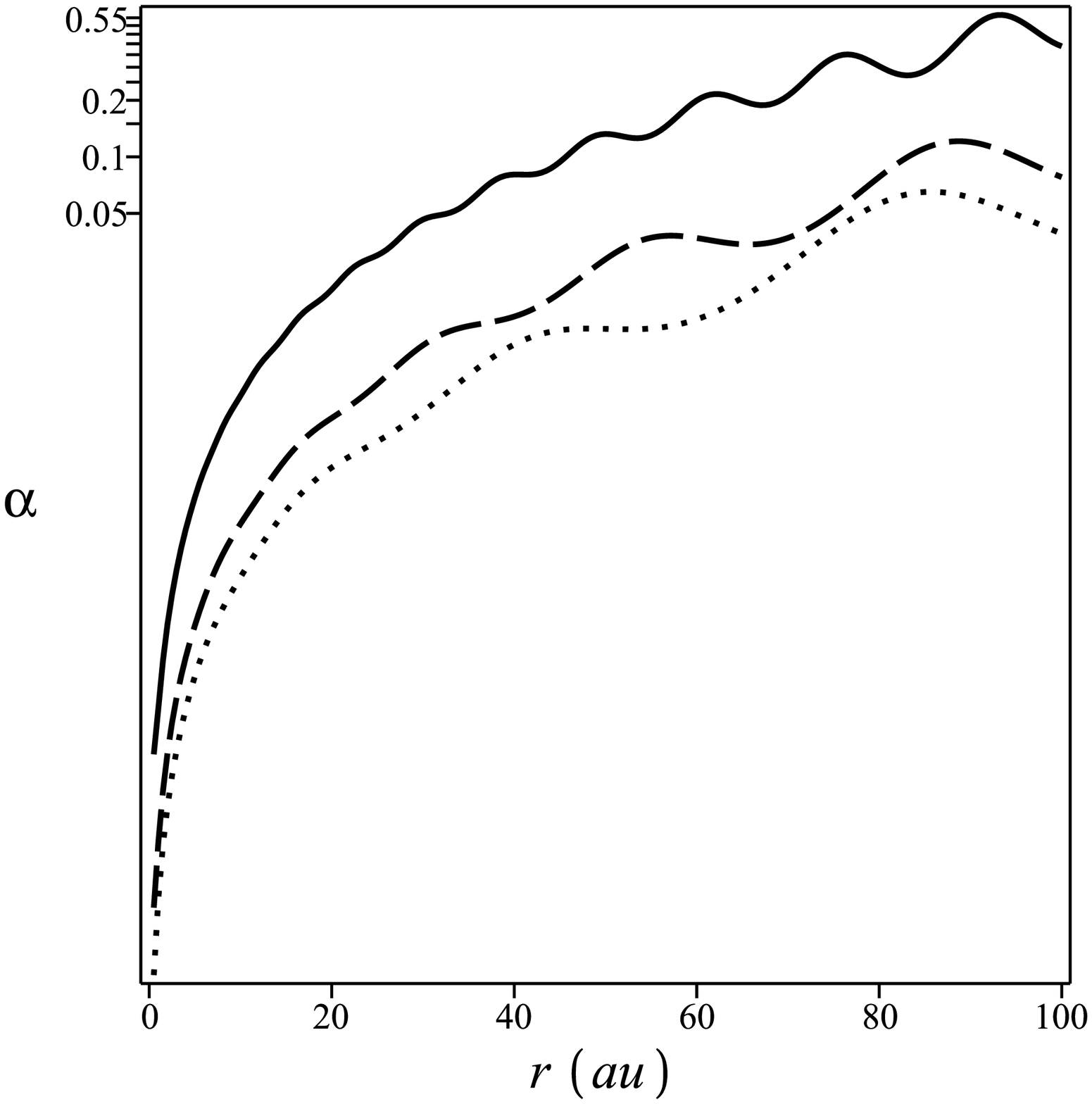}  }{\epsfxsize=5.0cm\epsffile{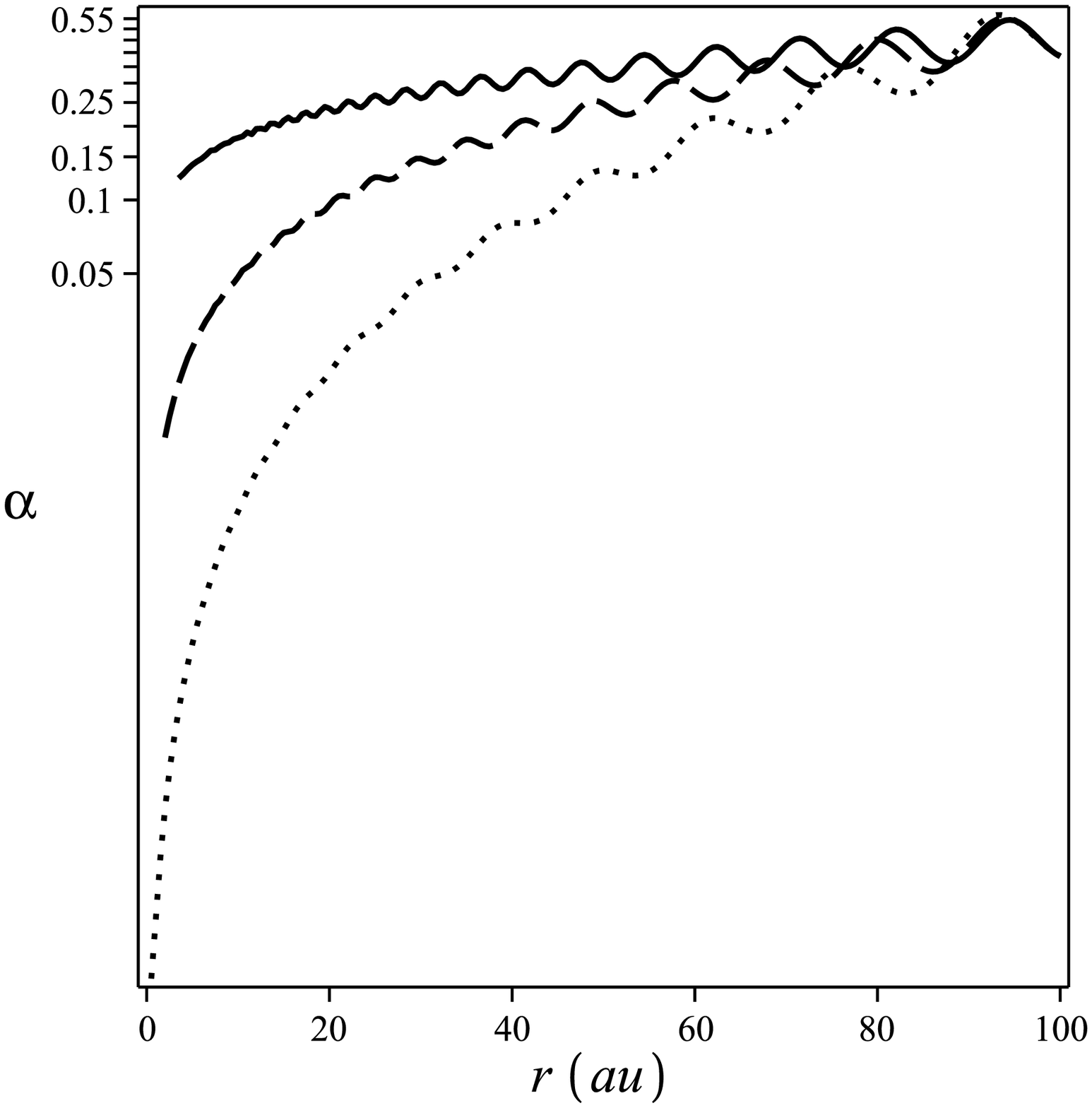}  }
} 
  \caption{The viscous parameter of $\alpha$ as a function of radius ($au$). The input parameters are set to the star mass $M_*=M_{\odot}$, the mass accretion rate  $\dot{M}=10^{-6} M_{\odot} yr^{-1}$, the ratio of the specific heats is set to be $\gamma=5/3$. \textit{Left panel} is for several values of Gammie's parameter $\beta_0$, the solid line represents $\beta_0=1$, the dashed line represents $\beta_0=5$, and the dotted line represents $\beta_0=10$, and $\delta=1.5$. \textit{Right panel} is for several values of parameter of $\delta$, the solid line represents $\delta=0.5$, the dashed line represents $\delta=1.0$, and the dotted line represents $\delta=1.5$, and $\beta_0=1.0$. 
}
\end{figure}

\subsection{The viscous parameter of $\alpha$ }                                            
In the present model, the viscous parameter of $\alpha$ depends on the physical quantities of the disc (Equation 10), especially the local cooling rate which depends on the local temperature. The profiles of the viscous parameter of $\alpha$ show that it increases by radii that this property is agree with simulation results of Rice \& Armitage (2009) and Rice et al. (2010). As, mentioned in the introduction, the minimum cooling timescale depends on the equation of state (Rice et al. 2005) with fragmentation occurring for $\tau_{cool} \leq 3 \Omega^{-1}$ when the specific heat ratio $\gamma=5/3$ (Gammie 2001). Rice et al. (2005) showed that fragmentation occurs for $\alpha > 0.06$ and this boundary is independent of the specific heat ratio $\gamma$. \textit{Left panel} of Fig 5 represents the viscous parameter of $\alpha$ as a function of radius for several values of the $\beta_0$ parameter. The solutions show the viscous $\alpha$ strongly depends on the $\beta_0$ parameter. As, the $\alpha$ parameter decreases by factor of $\beta_0$. Also, the solutions for small values of $\beta_0$ show the viscous $\alpha$ can reach to its critical value for fragmentation. \textit{Right panel} of Fig 5 represents the viscous parameter of $\alpha$ as a function of radius for several values of the $\delta$ parameter. The solutions represent the $\alpha$ parameter excluding the outer region of the disc strongly depends on the $\delta$ parameter. In $\delta=0.5$, the value of viscous $\alpha$ in whole of the disc is in the region for fragmentation. However, Rafikov (2005) suggested that it  is extremely difficult to see how fragmentation can occur within $10\, au$ even for the relatively massive discs. In $\delta=1.0$ and $\delta=1.5$, the viscous $\alpha$ in the inner disc ($r \lesssim 10$ and $40\, au$, respectively) is well below that required for fragmentation. 

The requirements for fragmentation are $Q \lesssim 1$ and $\alpha > 0.06$ (Rice et al. 2005, 2010; Rice \& Armitage 2009). In the present model, apparently the increase of the $\delta$ parameter reduces possibility of fragmentation (\textit{Right panel} of Fig 5). On the other hand, the increase of $\delta$ parameter can place the disc in gravitational instability (Fig 1). Thus, by a suitable value for the $\delta$ parameter, the disc can obtain two requirements for fragmentation. The Figs 1 and 5 imply that this value for small $\beta_0$ can be between $0.5$ and $1.0$.

\section{Summary and Discussion} 
In this paper, we have studied self-gravitating accretion discs in presence of a Newtonian potential of a point mass. We have used a prescription for cooling that is introduced by Gammie (2001). But, due to recent results of Cossins et al. (2010), we have assumed that cooling timescale in unit of dynamical timescale is a power-law function of temperature. Because of, the system equations are non-linear and there is not self-similar solution for it. First, we have obtained asymptotic solutions for system equations and then by them as boundary conditions, we integrated system equations numerically. 

The solutions showed that the structure of the disc strongly depends on the present cooling function. As, by adding importance degree of temperature in cooling timescale, gravitational instability extends from outer to inner radii. The solutions showed that in the case of cooling with temperature dependence, the disc thickness increases. But, this change of thickness is important in region with smaller Toomre parameter. In the present model, the effect of physical parameters studied such as mass accretion rate, $\beta_0$ parameter, and the ratio of the disc mass to central object mass. The results showed the structure of the disc is sensitive to these parameters. For example, the disc becomes gravitationally stable in larger mass accretion rate. While, the gravitational instability can occur in the larger disc mass. Also, the disc thickness increases by adding the mass accretion rate and decreases by adding the ratio of the disc mass to  the star mass. The study of the viscous parameter $\alpha$ in the present model shows that it increases by radii that this result is consistent with direct numerical simulations (e. g. Rice \& Armitage 2009; Rice et al. 2010). Also, the solution implies that the viscous $\alpha$ in the outer part of the disc becomes larger than its critical value ($\sim 0.06$) that might mean condition for fragmentation.

Here, the solutions represented that the disc thickness is very sensitive to input parameters. Thus, study of the present in a two dimensional approach may be interesting subject for future works. Also, it will be interesting to obtain a suitable $\delta$ value for fragmentation by direct numerical simulations. 

\section*{Acknowledgments}
I would like to acknowledge useful discussions with Alireza Khesali.

\end{document}